\documentclass[aps,prl,10pt,twocolumn,showpacs,superscriptaddress]{revtex4-1}
\usepackage{amsmath}
\usepackage{latexsym}
\usepackage{amssymb}
\usepackage{bm}
\usepackage{graphics,epstopdf}
\usepackage{color}

\usepackage{newlfont}
\usepackage{amsfonts}
\usepackage{amsthm}
\usepackage{graphicx}
\usepackage{epsfig}

\newcommand{\ket}[1]{|{#1}\rangle}
\newcommand{\bra}[1]{\langle{#1}|}

\newcommand{\expec}[1]{\langle#1\rangle}

\usepackage{times}

\newcommand{\be}{\begin{equation}}
\newcommand{\ee}{\end{equation}}
\newcommand{\bc}{\begin{center}}
\newcommand{\ec}{\end{center}}
\newcommand{\bea}{\begin{eqnarray}}
\newcommand{\eea}{\end{eqnarray}}
\newcommand{\ba}{\begin{array}}
\newcommand{\ea}{\end{array}}

\begin{document}
\title{Simultaneous polarization squeezing in polarized $ N $ photon state and diminution on a squeezing operation}

\author{Namrata Shukla}
\email{namratashukla@hri.res.in}

\affiliation{Now at, Quantum Information and Computation Group,\\
Harish-Chandra Research Institute, Chhatnag Road, Jhunsi,
Allahabad 211 019, India}

\affiliation{Department of Physics,\\
University of Allahabad, Allahabad, Allahabad-211001, UP, India}



\author{Ranjana Prakash}


\affiliation{Department of Physics,\\
University of Allahabad, Allahabad, Allahabad-211001, UP, India}



\date{\today}

\begin{abstract}
We study polarization squeezing of a pure photon number state 
which is obviously polarized but the mere change in the basis of 
polarization leads to simultaneous polarization squeezing 
in all the components of Stokes operator vector except those falling 
along or perpendicular to the direction of polarization state, is observed. 
We use the most general definition of polarization squeezing and discuss the experimental 
feasibility of the result. We also observe that a squeezing operation like non-degenerate 
parametric amplification of the state does not reveal simultaneous squeezing in all 
Stokes operator vectors and decreases in this sense.
\end{abstract}

\maketitle

\section{Introduction}
\label{intro}
Products of quantum fluctuations in two non commuting observables satisfy the uncertainty relation 
but the individual fluctuations can be reduced and this gives the well known concept of squeezing. 
In classical optics, the polarization state of light  beam can be visualized as direction of 
a Stokes vector in poincare sphere and is determined by the four Stokes 
parameters \cite{Stokes-para,Bo-wolf}. For a monochromatic unidirectional light 
traveling along z-direction, the classical Stokes parameters $S_{0}$ and 
${\bm S}=S_{1}, S_{2},S_{3}$ are defined as
\begin{equation}
\label{eq1}
S_{0,1}= \expec{\bm{\mathcal{E}}_{x}\bm{\mathcal{E}}_{x}^{*}}
\pm\expec{\bm{\mathcal{E}}_{y}\bm{\mathcal{E}}_{y}^{*}}~,~
S_{2}+i S_{3}=2 \expec{\bm{\mathcal{E}}_{x}^{*}\bm{\mathcal{E}}_{y}},
\end{equation}
where $ {\bm{\mathcal E}_{x, y}}$ are the components of analytic signal 
for the electric field. For perfectly polarized light
\begin{equation}
\label{eq2}
 S_{0}^2=S_{1}^2+S_{2}^2+S_{3}^2,
\end{equation}
and the point $(S_{1}, S_{2}, S_{3})$ is on a sphere of radius $S_{0}$, called the 
Poincare sphere. \\

Quantum mechanical analogue of Stokes parameters can also be defined to characterize quantum nature 
of polarization. These are observables and can be associated with hermitian operators $ \hat{\bm S}=\hat S_{1,2, 3}$ and defined as
\cite{Man-wolf},
\begin {equation}
\label{eq3}
 \hat S_{0, 1}=\hat a_{x}^\dagger \hat a_{x}\pm \hat a_{y}^\dagger \hat a_{y},~
 \hat S_{2}+i \hat S_{3}=2 \hat a_{x}^\dagger \hat a_{y},
\end {equation}
 where $ \hat a_{x, y}$ and $ \hat a_{x, y}^\dagger$ are annihilation 
 and creation operators, respectively for the two orthogonal linear polarization modes $x$ and $y$. 
 These Stokes operators obey the commutation relations 
 \begin {equation}
 \label{eq4} 
 [\hat S_0,\hat S_j]=0,~[\hat S_j,\hat S_k]=2i\sum_{l}\epsilon_{jkl} ~\hat S_l,
 \end {equation}
following the SU(2) algebra. Here, $\epsilon_{jkl}$ is Levi-Civita symbol for $(j,k,l=1,2,$ or $3)$. 
The obvious uncertainty relations for the fluctuation in Stokes operators are,
 \begin{equation}
 \label{eq5}
  V_{j} V_{k}\geqslant{\expec{\hat S_{l}}}^2,~ 
  V_{j}\equiv\expec{\hat S_{j}^2}- {\expec{\hat S_{j}}}^2.
 \end{equation} 
 
Polarization of optical fields is a well known and old concept in classical optics and 
there is a very good correspondence between Stokes parameters and Stokes operators. The Stokes parameters 
involve coherence functions \cite{Man-wolf} of order (1,1) and it
has been realized that these are insufficient to describe polarization completely 
and, {\it e.g.}, $\bm{S}=0$ does not represent only unpolarized light \cite{HP-NC}. These parameters
still remain important mainly due to the non-classicalities associated with polarization, {\it viz}, 
polarization squeezing \cite{Pol-sq 1,Pol-sq 2,Pol-sq 3,Pol-sq 4,Pol-sq 5,Pol-sq A} and 
polarization entanglement \cite{Pol-eng}.\\

The Stokes operators are the relevant continuous variables for the system 
describing polarization. Squeezed radiation states in quantum optics are
identified by the property that their quantum fluctuations are reduced below the standard quantum limit in
one of the quadrature components. Similar to this concept, polarization squeezing is defined
using the commutation relations followed by Stokes operators and uncertainty products. 
Existence of a minimum quantum limit for product of uncertainties in measurement 
of two different Stokes operators leads to the concept of polarization squeezing discussed 
in the next section, where one of the uncertainties $ V_{j}$ and $ V_{k}$ can be reduced 
below $ \mid\expec{\hat S_{l}}\mid$ at the expense of the other, {\it e.g.} in \eqref{eq5}. \\

These properties are of paramount importance in the real world applications 
on different scales as quantum Stokes operators and non-classical polarization can be used for quantum 
information protocols and quantum communication. It is convenient to study polarization squeezing because 
it is easy to measure the stokes parameters using linear optical elements and thus polarization squeezing is easy to 
experimentally observe. The direct measurement schemes for measuring these parameters are developed
methods which preserve quantum noise property. Looking forward to such applications, it is 
desirable to devise methods for generation of states with appreciable polarization squeezing.\\

In the present paper, we first discuss the significance of different criteria for polarization squeezing 
in the next section. In section 3, We use the most general criterion \cite{def-me} for polarization 
squeezing to show that in case of a polarized $ N $ photon number state, all the three orthogonal components of 
Stokes operator vector may be squeezed and as a matter of fact, almost all components are squeezed. We consider 
the polarization squeezing along a general component of Stokes operator vector and show that there is polarization squeezing 
unless the component is either along or perpendicular to the direction of polarization of light.
In section 4, we show that after a squeezing operation like parametric amplification applied to a photon number state, squeezing 
is observed only in certain Stokes operators in specific conditions.

\section{Polarization squeezing}
\label{sec:1}
First definition of polarization squeezing is due to Chirkin et al. \cite{Pol-sq 1} in terms of
variances of Stokes operators for a given state and for an equally intense coherent state, written as
\begin {equation}
\label{eq6}
 V_{j}<V_{j} (coh)=\expec{\hat S_{0}},
\end {equation}
, {\it i.e.}, $\hat S_{j}$ is squeezed if $ V_{j}$ is less than $ V_{j}$ for equally intense 
coherent light which gives the value $ \expec{\hat S_{0}}$ of the variance.
This definition has been used by some authors \cite{Pol-sq 2}.\\

Heersink et al. \cite{Pol-sq 3} defined polarization squeezing using the uncertainty relations
\eqref{eq5} in the form
\begin {equation}
\label{eq7}
 V_{j}<|\expec{\hat S_{l}}|<V_{k}, ~ j\neq k \neq l,
\end {equation}
for squeezing of $ \hat S_{j}$. This definition has also been used by 
some authors \cite{Pol-sq 4}.\\

Luis \cite{Pol-sq 5} considered various criteria for polarization
squeezing and compared their stringency. He finally gave the criterion for polarization squeezing
of a component of $\hat {\bm S}$ along a unit vector ${\bm n}$ as
\begin{equation}
\label{eq8}
 V_{\bm n}<|\expec{\hat S_{\bm n_{\perp}}}|,
\end{equation}
here $ \hat S_{\bm n_{\perp}}$ is component of $\hat {\bm S}$ along unit vector $ \bm n_\perp $ 
which is perpendicular to $ \bm n $. For suitable orthogonal components $ \bm n $ and 
$ \bm n_{\perp}$, he discussed the order of stringency of the various criteria and it can be represented as
\begin{equation}
\label{eq9}
 V_{\bm n}<\expec{\hat S_{\bm n_{\perp}}}^2 / \expec{\hat S_{0}}
 <|\expec{\hat S_{\bm n_{\perp}}}|<\expec{\hat S_0}.
\end{equation}\\

The authors finally have written the criterion for polarization squeezing \cite{def-me} in 
the form 
\abovedisplayskip
\abovedisplayskip
\begin{eqnarray}
\label{eq10}
V_{\bm n}\equiv\expec{\Delta \hat S_{\bm n}^2} &&< {\mid\expec{\hat S_{\bm n_{\perp}}}\mid}_{max}\nonumber\\
&&=\sqrt{{\mid\expec{\hat {\bm S}}\mid}^2-{\expec{\hat S_{\bm n}}}^2},
\end{eqnarray}
arguing that for a given component $ \hat S_{\bm n} $ there are infinite directions 
$ \bm n_\perp $ and consideration of the maximum possible value of 
$ |\expec{\hat S_{\bm n_{\perp}}}| $. \\

It can be seen that each of Eqs. \eqref{eq6} - \eqref{eq8} and \eqref{eq10} describes a non-classicality
in quantum optics but only \eqref{eq7}, \eqref{eq8} and \eqref{eq10} are related to uncertainty relations. 
We shall use the criterion \eqref{eq10} for polarization squeezing which is most general and based on the actual
uncertainty relations. We may also define squeezing factor $ \mathcal{S}_{\bm n} $ and degree of squeezing 
$ \mathcal{D}_{\bm n} $ by writing
\begin{equation}
\label{eq11}
 \mathcal{S}_{\bm n}=\frac{V_{\bm n}}{\sqrt{{|\expec{\hat {\bm S}}|}^2-{\expec{\hat S_{\bm n}}}^2}},~ 
 \mathcal{D}_{\bm n}=1-\mathcal{S}_{\bm n}.
\end{equation}
Non-classicalities appear when $ 1>\mathcal{S}_{\bm n}>0 $ and the degree of squeezing $ \mathcal{D}_{\bm n}$
lies between $0$ and $1$. If $\rho$ is density operator of radiation given in the diagonal Sudarshan-Glauber representation 
\cite{Sudarshan-glauber} by
\begin{equation}
\label{eq12}
 \rho=\int{ d^2\alpha~d^2\beta~P(\alpha, \beta) \ket{\alpha, \beta} \bra{\alpha, \beta}}.
\end{equation}
where $\ket{\alpha, \beta}$ are the coherent states defined as
$ (a_{x}, a_{y})\ket{\alpha, \beta}=( \alpha, \beta)\ket{\alpha, \beta}$. Eq. \eqref{eq6}
then gives
\begin{equation}
\label{eq13}
 V_{j}-\expec{\hat S_{0}}=\int d^2\alpha~d^2\beta~P(\alpha, \beta)~[f_j(\alpha, \beta)-\expec{f_{j}}]^2,\\
\end{equation}
with
\begin{eqnarray}
\expec{f_j}=\int d^2\alpha~d^2\beta~P(\alpha, \beta)~[f_j(\alpha, \beta)], \nonumber\\
f_1(\alpha, \beta)\equiv {|{\alpha}|}^2-{|{\beta}|}^2,\nonumber
f_2+if_3= 2{\alpha \beta},
\end{eqnarray}
while $ f_{j}'s $ are real functions of $ \alpha $ and $ \beta $ and  holding of 
$ V_{j}<\expec{\hat S_{0}} $ rules out possibility of existence of a non-negative weight function 
$ P(\alpha, \beta) $ which could be identified to a classical probability distribution. Similarly, Eq. 
\eqref{eq10} for $ j=1 $ gives

\begin{eqnarray}
\label{eq15}
&&V_{1}-\sqrt{{|\expec{\hat {\bm S}}|}^2-{\expec{\hat S_{1}}}^2}\nonumber\\
&&=|\expec{\hat {S_1}^2}|-{|\expec{\hat {S_1}}|}^2-|\expec{\hat S_2+i\hat S_3}|\nonumber \\
&&=\int d^2\alpha~d^2\beta~P(\alpha, \beta)~[(f_1(\alpha, \beta)-\expec{f_1})^2+(|\alpha|-|\beta|)^2]\nonumber\\
&&<0.
\end{eqnarray}
which also denies the existence of a positive definite $ P(\alpha, \beta) $.
It can be seen that if Eq.\eqref{eq15} which represents criterion \eqref{eq10} holds then Eq.
\eqref{eq13} representing criterion \eqref{eq6} has to hold. 

\section{Polarization squeezing of Photon number state}
\label{sec:2}
Let us now consider a polarized $ N $ photon state $ \ket \psi $ traveling along z-direction, 
the polarization being given by the unit complex vector
\begin{equation}
\label{eq16}
\bm\varepsilon= \varepsilon_x \bm e_x+\varepsilon_y \bm e_y,~ \varepsilon_x=\cos{\frac{\theta}{2}}, ~ \varepsilon_y=e^{i\phi} \sin{\frac{\theta}{2}},
\end{equation}
where $ \bm e_{x,y} $ are unit vectors along the x and y-directions.
On the Poincare sphere, this polarization state can be represented by the unit vector
\begin{equation}
\label{eq17}
\bm m =\cos\theta~ \bm e_{x} + \sin\theta(\cos\phi~ \bm e_y+\sin\phi~ \bm e_z).
\end{equation}
Mode orthogonally polarized to $ \bm \varepsilon $ can be represented by another unit polarization vector
\begin{equation}
\label{eq18}
\bm\varepsilon_\perp = \varepsilon_{\perp x}\bm e_{x}+ \varepsilon_{\perp y}\bm e_{y},~
\varepsilon_{\perp x}=-\sin{\frac{\theta}{2}}, \varepsilon_{\perp y}= e^{i\phi}\cos{\frac{\theta}{2}}.
\end{equation}
Annihilation operators for light in this changed basis $(\bm \varepsilon, \bm\varepsilon_\perp)$ are given by
\begin{equation}
\label{eq19}
 \hat a_{\bm \varepsilon}=\varepsilon_{x}^* \hat a_{x} + \varepsilon_{y}^* \hat a_{y},~
 \hat a_{\bm \varepsilon_{\perp}}=\varepsilon_{\perp x}^* \hat a_{x} + \varepsilon_{\perp y}^* \hat a_{y},
\end{equation}
which allows us to write the same in $(x, y)$ mode as
\begin{equation}
\label{eq20}
 \hat a_x=\varepsilon_{x} \hat a_{\bm \varepsilon}+\varepsilon_{\perp x} \hat a_{\varepsilon_{\perp}},~
 \hat a_y=\varepsilon_{y} \hat a_{\bm \varepsilon}+\varepsilon_{\perp y} \hat a_{\varepsilon_{\perp}}.
\end{equation}
The state $ \ket \psi $ in $ (\bm \varepsilon, \bm\varepsilon_\perp)$ can then be written as
\begin{equation}
\label{eq21}
\ket \psi=(N!)^{1/2}{(\hat a_{\bm\varepsilon}^\dagger)}^N \ket {vac},
\end{equation}
where $\ket{vac} $ is vacuum state satisfying 
$ \hat a_{\bm \varepsilon} \ket{vac} = 0$. On considering the normal ordering of the operators \cite{def-me}, expectation values of 
Stokes operators \eqref{eq3} and their squares and anti-commutators can be 
obtained by straight calculations which on simplification (using $ \hat a_{\bm\varepsilon_{\perp}}\ket\psi=0 $)
give
\begin{eqnarray}
\label{eq22}
&&\expec{\hat S_0}=N,~ \expec{\hat S_1}=N \cos{\theta},\nonumber \\
&&\expec{\hat S_2}=N sin\theta \cos\phi,~ \expec{\hat S_3}=N \sin\theta sin\phi, \nonumber \\
&&\expec{\hat S_{0}^2}=N(N-1),\nonumber\\
&&\expec{\hat S_{1}^2}=N(N-1) \cos^2 \theta + N,\nonumber \\
&&\expec{\hat S_{2}^2}=N(N-1)\sin^2 \theta \cos^2 \phi + N,\nonumber \\ 
&&\expec{\hat S_{3}^2}=N(N-1) \sin^2 \theta \sin^2 \phi,\nonumber \\
&&\expec{\{\hat S_1 , \hat S_2\}}=2N(N-1)\cos\theta \sin\theta \cos\phi, \nonumber \\
&&\expec{\{\hat S_1 , \hat S_3\}}=2N(N-1)\cos\theta \sin \theta \sin\phi, \nonumber \\
&&\expec{\{\hat S_2 , \hat S_3\}}=2N(N-1)\sin\theta \cos\phi \sin\phi.
\end{eqnarray}
Now, to study squeezing of component $ \hat S_{\bm n}\equiv\bm n.{\hat {\bm S}} $ of Stokes 
vector along unit vector $\bm n=(n_1,n_2,n_3)$, we can write down the general expressions

\begin{equation}
\label{eq23}
\expec{\hat S_{\bm n}}=N(\bm n .\bm m), \expec{\hat S_{\bm n}^2}=N(N-1){(\bm n.\bm m)}^2 +N.
\end{equation}
The variance $ V_{\bm n} $ of this general component $ \hat S_{\bm n}$ is
\begin{equation}
\label{eq24}
 V_{\bm n}=N[1-{(\bm n.\bm m)}^2],
\end{equation}
and for polarization squeezing this is to be compared with maximum value of 
modulus of expectation value of component of $ \hat{\bm S} $ perpendicular to $\bm n $, {\it i.e.}, with
\begin{eqnarray}
\label{eq25}
|\expec{\hat S_{\bm n_{\perp}}}|_{max}&&= {\big({|\expec{\hat {\bm S}}}^2|-{\expec{\hat S_{\bm n}}}^2\big)}^{1/2}\nonumber\\
&&=N{[1-{(\bm n.\bm m)}^2]}^{1/2}.
\end{eqnarray}
Since, $ 1-{(\bm n.\bm m)}^2<1 $ for $(\bm n.\bm m)\neq0 $, Eq. \eqref{eq24} and \eqref{eq25}
make it clear that, $ V_{\bm n}<|\expec{\hat S_{\bm n\perp}}| $ holds unless
$(\bm n.\bm m)=0,1 $ with squeezing factor and degree of squeezing is
\begin{eqnarray}
\label{eq26}
&&\mathcal{S}_{\bm n}={[1-{(\bm n.\bm m)}^2]}^{1/2},\nonumber\\
&&\mathcal{D}_{\bm n}=1-\mathcal{S}_{\bm n}=1-{[1-{(\bm n.\bm m)}^2]}^{1/2}.
\end{eqnarray}
We observe, $ \mathcal{S}_{\bm n}<1 $ for all $ \bm n $ and therefore all components $ \hat S_{\bm n} $ are squeezed unless $ \bm n $ is 
along or perpendicular to $ \bm m $. Thus, for any polarized $ N $ photon state, inequalities giving polarization squeezing 
are satisfied for all components of Stokes operators and squeezing occurs  
unless the component is along a direction which is perpendicular or same as the direction describing 
polarization of light on Poincare sphere. \\

Regarding the experimental observation of this effect, it should be noted that for 
$ \bm n = (\cos\theta_0,~\sin\theta_0 cos\phi_0,~sin\theta_0 sin\phi_0)$, Eq. \eqref{eq19}
gives,
\begin{equation}
\begin{split}
\bm n.\hat{\bm S}&=\cos\theta_0(\hat a_{x}^\dagger \hat a_x-\hat a_{y}^\dagger \hat a_y)
+\sin\theta_0(e^{-i\phi_0} \hat a_{x}^\dagger \hat a_y + e^{i\phi_0} \hat a_{y}^\dagger \hat a_x)\\
 &=\hat a_{\varepsilon_0}^\dagger \hat a_{\varepsilon_0}-\hat a_{\varepsilon_{0\perp}}^\dagger \hat a_{\epsilon_{0\perp}},\nonumber
\end{split}
\end{equation}
where 
\begin{eqnarray}
&&\bm \varepsilon_0=\cos\frac{\theta_0}{2} \bm e_x + e^{i\phi_0} \sin\frac{\theta_0}{2}\bm e_y, \nonumber\\
&&\bm\varepsilon_{0\perp}=-\sin\frac{\theta_0}{2} \bm e_x +e^{i\phi_0} \cos\frac{\theta_0}{2} \bm e_y.\nonumber
\end{eqnarray}
Thus, to measure $\expec{\bm n.{\hat {\bm S}}}$ and $ {\expec{\bm n.{\hat {\bm S}}}}^2 $
one has to make measurements of $ (\hat N_{\varepsilon_0}- \hat N_{\varepsilon_{0\perp}}) $ and  
$ {(\hat N_{\varepsilon_0}- \hat N_{\varepsilon_{0\perp}})}^2 $, where $ \hat N_{\varepsilon_0} $
and $ \hat N_{\varepsilon_{0\perp}} $ are the photon number operators 
$ \hat a_{\varepsilon_0}^\dagger \hat a_{\varepsilon_0} $ and 
$ \hat a_{\varepsilon_{0\perp}}^\dagger \hat a_{\varepsilon_{0\perp}} $, respectively
and this can be done easily by
\begin{enumerate}
\item Introducing phase shift $ \phi_0 $ in the $y$-linearly polarized mode.
\item Rotating the plane of polarization by angle $\frac{\theta_0}{2}$.
\item Measuring $(\hat N_x-\hat N_y)$ and ${(\hat N_x-\hat N_y)}^2$ in the changed basis. 
\end{enumerate}

\section{Photon number state under squeezing operation}

Photon number state polarized in mode $ (\bm \varepsilon,\bm\varepsilon_\perp) $ with no photon
in polarization mode $ \bm\varepsilon_\perp $ given by Eq. \eqref{eq21} is now subjected to some nonlinear 
interaction like parametric amplification \cite{Par-amp}. The interaction hamiltonian for such an operation
is $H=g\big(\hat a_{x}^\dagger\hat a_{y}^\dagger+\hat a_{x}\hat a_{y}\big)$ and the annihilation and 
creation operator after the interaction for time $ t $ can be written as
\begin{eqnarray}
\label{eq27}
 &&\hat a_{x}(t)=(\cosh gt)\hat a_{x}-i(\sinh gt)\hat a_{y}^\dagger, \nonumber\\
 &&\hat a_{y}(t)=(\cosh gt)\hat a_{y}-i(\sinh gt)\hat a_{x}^\dagger,
\end{eqnarray}
where $ \hat a_x $ and $ \hat a_y $ given by Eq. \eqref{eq20} with 
$ (\varepsilon_x, \varepsilon_{\perp x})$ and $ (\varepsilon_y, \varepsilon_{\perp y})$ 
written in Eqs. \eqref{eq16} and \eqref{eq18}.
Average values of Stokes parameters and their variances are obtained as
\begin{eqnarray}
\label{eq28}
&&\expec{\hat S_1}=N \cos {\theta}, \nonumber\\
&&\expec{\hat S_2}=N(c^2+s^2) \sin\theta \cos\phi,\nonumber\\
&&\expec{\hat S_3}=N(c^2+s^2) \sin \theta \sin \phi,\nonumber\\
\end{eqnarray}
and
\begin{eqnarray}
V_1&=&N{\sin}^2\theta,\nonumber\\
V_2&=&N(c^2+s^2)(1-{\sin}^2\theta {\cos}^2\phi)\nonumber\\
&&+2c^2 s^2 [N^2(1-{\sin}^2\theta {\sin}^2\phi)+N(1+{\sin}^2\theta {\sin}^2\phi)+2],\nonumber\\
V_3&=&N(c^2+s^2)(1-{\sin}^2\theta {\sin}^2\phi)\nonumber\\
&&+2c^2 s^2 [N^2(1-{\sin}^2\theta {\cos}^2\phi)+N(1+{\sin}^2\theta {\cos}^2\phi)+2],\nonumber\\
\label{eq29}
\end{eqnarray}
where $ c=\cosh gt $ and $ s=\sinh gt $, $ gt $ being interaction time. \\

As, it is clear by having a look at the above expressions for mean values and variances 
of Stokes operators, it is not trivial to generalize these expressions for the Stokes operator 
$ \hat S_{\bm n} $ along unit vector $ \bm n $. But, it is interesting to find the extent of 
squeezing along the three Stokes vectors $ \hat S_1, \hat S_2 $ and $ \hat S_3 $, fitting these 
values in the criterion \eqref{eq11}.\\

If we see squeezing in $ \hat S_1 $ operator, the squeezing factor on substituting the 
appropriate values and simplifying the expression is
\begin{equation}
\label{eq30}
\mathcal{S}_1=\frac{N {\sin}^2\theta}{\sqrt{N^2{(c^2+s^2)}^2 {\sin}^2\theta}}
=\frac{sin\theta}{\cosh2gt},
\end{equation}
which shows squeezing in $ \hat S_1 $ component as $ \sin\theta<\cosh2gt $. 
Degree of squeezing is thus given by
\begin{equation}
\mathcal{D}_1=1-\frac{\sin\theta}{\cosh2gt}.\nonumber 
\end{equation}
We may compare the degree of squeezing in this case with $ \mathcal D_1=1-\sin\theta $,~ 
at $ gt=0 $. It shows that the squeezing in $ \hat S_1 $ increases after the parametric 
amplification like operation on the considered state. \\

\begin{figure}
\centering
\includegraphics[width=150pt]{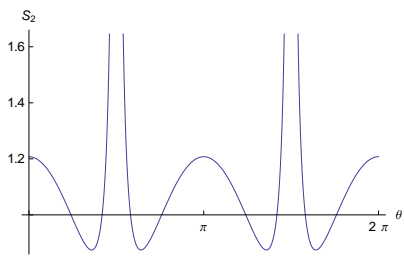}
\caption{Variation of squeezing factor $\mathcal{S}_2$ with $\theta$ for $N=8,~gt=0.1$ for $\phi=0$.}
\label{a}
\end{figure}

\begin{figure}
\centering
\includegraphics[width=150pt]{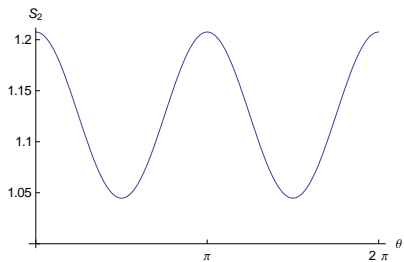}
\caption{Variation of squeezing factor $\mathcal{S}_2$ with $\theta$ for $N=8,~gt=0.1$ for $\phi=\pi/2$.}
\label{b}
\end{figure}

To investigate squeezing in the component $ \hat S_2 $, the expression for squeezing 
factor can be written as
\begin{eqnarray}
\label{eq31}
\mathcal{S}_2=\frac{N(c^2+s^2)(1-{\sin}^2\theta {\cos}^2\phi)}{\sqrt{N^2 {\cos}^2\theta
+N^2{(c^2+s^2)}^2 {\sin}^2\theta {\sin}^2\phi}}\nonumber\\
+\frac{2c^2 s^2[N^2+N+2-
(N^2-N){\sin}^2\theta {\sin}^2\phi]}{\sqrt{N^2 {\cos}^2\theta
+N^2{(c^2+s^2)}^2 {\sin}^2\theta {\sin}^2\phi}}.
\end{eqnarray}
To have a better insight of the squeezing along $\hat S_2$, we plot this expression with respect to $\theta$
for a fixed number of photons $N=8$. Fig. \ref{a} and Fig. \ref{b} clearly indicate the occurrence 
of polarization squeezing in the following two cases derived from Eq. \eqref{eq31}. 
\begin{itemize}
\item [Case 1]:
For plane polarized light, {\it i.e.}, $ \sin\phi=0 $,
\begin{equation}
\label{eq32}
\mathcal{S}_{2}={(c^2+s^2)}^2 |\cos\theta|+2c^2 s^2 
\bigg[\frac{N^2+N+2}{N}\bigg]|\sec\theta|.
\end{equation}
This expression is arithmetic mean of the double of the two terms which gives 
the minimum value of $ \mathcal{S}_2 $ as double of the geometric mean of the two terms as
\begin{equation}
\mathcal{S}_{2}(min)=2\sqrt{2{(c^2+s^2)}^2 \bigg[\frac{N^2+N+2}{N}\bigg]c^2 s^2},\nonumber
\end{equation}
and given after simplification as
\begin{equation}
\label{eq33}
\mathcal{S}_{2}(min)=2\sqrt{\bigg[\frac{N^2+N+2}{8N}\bigg]}\sinh4gt.
\end{equation}
This expression shows squeezing in polarization for very small times of interaction and
\begin{equation}
\cos\theta=\tanh2gt\sqrt{\bigg[\frac{N^2+N+2}{N}\bigg]}.\nonumber
\end{equation}
\item [Case 2]: 
For circularly polarized light, {\it i.e.}, $ \sin\theta=1$ and $ \cos\phi=0 $, 
squeezing factor in this case observed after simplification is
\begin{equation}
\label{eq34}
\mathcal{S}_2=\cosh2gt+{\bigg[\frac{2N+2}{N}\bigg]}{\frac{{\sinh}^22gt}{\cosh2gt}},
\end{equation}
and it seems to be greater than unity resulting in no squeezing for circularly polarized light.
\end{itemize}
For fixed number of photons $N$, $\phi=0$ shows that polarization squeezing may occur for very small interaction times, {\it e.g.}, $gt=0.1$ as seen in
Fig.~\ref{a} but there is no squeezing for $\phi=\pi/2$ as depicted in Fig.~\ref{b}.\\

We also check the polarization squeezing along $\hat S_3 $ however it does not reveal squeezing in either of the cases 
of circularly or plane polarized light.

  
\section{Result and Discussion}
We found the simultaneous polarization squeezing in almost all components of Stokes operator vector except those 
along or perpendicular to the direction of polarization just by changing the basis of polarization.
Most importantly, it happens under no condition and gives a general result about the photon number state. \\

However, we observe that the simultaneous polarization squeezing of all the components of Stokes operator 
in pure polarized photon number state is not observed after a squeezing operation. 
However, the state, after it is subjected to a nonlinear interaction for time $t$ is found 
to be more squeezed than at $gt=0$ along $ \hat S_1 $ irrespective of the nature of 
polarization. Squeezing along $ \hat S_2 $ is seen for plane polarized light, however,
$\hat S_3 $ does not exhibit any squeezing. This result is in agreement with the results 
reported in the previous investigation
which shows simultaneous polarization squeezing in almost all components
of Stokes operator vector $ \hat{\bm S} $ except for which
$ \bm n.\bm m=0 $ or $ 1 $, where $\bm m $ 
represents polarization of pure photon number state in Poincare sphere. \\

The non-occurrence of squeezing in $\hat S_2 $ at $ gt=0 $ for circularly
polarized light is clear from the fact that in this case, $\bm m $ is along or 
opposite to z-axis and is perpendicular to $ \bm n $ which is along $\hat S_2 $. 
Similarly, non-occurrence of squeezing in $\hat S_3 $ at $ gt=0 $ for plane polarized light
is evident as $\bm n $ is along z- axis and $\bm m $ in x-y plane, while in case of 
circular polarization $ \bm n $ is along z-direction and $\bm m $ is either along z-direction
or opposite to it. The expressions for squeezing factor for $ \hat S_2 $ as a function of interaction time 
shows the occurrence of polarization squeezing for very small interaction times in case of plane polarized light. 
The pattern of polarization squeezing with the angles $\theta$ and $\phi$ describing polarization is 
shown in Fig.~\ref{a} and Fig.~\ref{b}. \\

\section{Acknowledgement}
We would like to thank Hari Prakash for his interest and critical comments.

\end{document}